\documentclass[12pt]{article}
\usepackage{amsmath}
\usepackage{amsthm}
\usepackage{geometry}
\usepackage{fancyhdr}
\usepackage{listings}
\usepackage{xcolor}
\usepackage{amssymb}
\usepackage{graphicx}
\usepackage{amsfonts}
\usepackage{multirow}
\usepackage{diagbox}
\usepackage{array}
\usepackage{booktabs}
\usepackage{mathrsfs}
\usepackage{subfigure}
\usepackage{setspace}
\usepackage{threeparttable}
\usepackage[round]{natbib}
\usepackage[colorlinks=true, allcolors=black]{hyperref}%
\usepackage{makecell}
\usepackage[figuresright]{rotating}
\usepackage[ruled,linesnumbered]{algorithm2e}
\usepackage{fullpage}
\theoremstyle{plain}
\newtheorem{theorem}{Theorem}
\newtheorem{assumption}{Assumption}

\theoremstyle{remark}
\newtheorem{remark}{Remark}[section]

\def\E{\mathrm{E}}
\def\Var{\mathrm{Var}}
\def\Cov{\mathrm{Cov}}
\def\T{\mathrm{T}}

\newcommand\blfootnote[1]{%
  \begingroup
  \renewcommand\thefootnote{}\footnote{#1}%
  \addtocounter{footnote}{-1}%
  \endgroup
}

\title{Estimating Treatment and Spillover Effects with the Ego-Cluster Experimental Design}
\date{}
\author{%
{Xiao Liu$^{1}$, Feifang Hu$^{2}$, and Jingfei Zhang$^1$}
\vspace{1.6mm}\\
\fontsize{11}{10}\selectfont\itshape $^1$\,Goizueta Business School, Emory University, Atlanta, GA, USA. \\
\fontsize{11}{10}\selectfont\itshape $^2$\,Department of Statistics, George Washington University, DC, USA.}

\usepackage{hyperref}
\begin{document}
\maketitle

\blfootnote{Address for correspondence: Jingfei Zhang, Goizueta Business School, Emory University, Atlanta, GA, USA. \url{emma.zhang@emory.edu}.}

\begin{abstract}
\baselineskip=20pt
Network interference occurs when a unit’s outcome depends not only on its own treatment but also on the treatments received by connected units in the network. Experimental designs and analysis methods that ignore such interference can yield biased estimators of causal effects. In this paper, we develop a new experimental design for {the estimation and inference} of global treatment effect and spillover effect under a model-based framework and ego-cluster randomization. Under this design, the network is partitioned into a collection of ego-clusters, each consisting of a focal unit (the ego) and its network neighbors (the alters), with randomization conducted at the cluster level. We propose model-based estimators for the global treatment effect and spillover effect and establish their consistency and asymptotic normality, with asymptotic variances determined by the ego-cluster structure. Building on these theoretical results, we introduce an ego-clustering algorithm that sequentially selects egos and assigns alters to minimize asymptotic variances. Simulation studies and two empirical applications demonstrate that the proposed procedure yields accurate inference and efficiency improvements over existing network experimental designs.

\vspace{0.2in}
\end{abstract}

\newpage
\baselineskip=26.5pt

\section{Introduction}\label{sec:intro}
Randomized controlled experiments provide a foundational tool for studying causal effects. A key assumption in classical experimental designs and analysis methods is the \textit{Stable Unit Treatment Value Assumption} (SUTVA) \citep{Rubin2015}, which states that each unit's outcome depends only on its own treatment and is unaffected by treatments assigned to others. However, in experiments where subjects are connected through a network, the SUTVA assumption is often violated, as a unit's outcome depends not only on its own treatment but also on the treatments received by connected units in the network \citep{Manski2000economic}. Such network interference is common in experiments in social networks, education, finance, and economics \citep{banerjee2013diffusion, Phan2015, Paluck2016}. The violation of SUTVA poses challenges for the design and analysis of network experiments, as classical methods are no longer valid under network interference \citep{manski2013identification}.

A growing methodological literature studies causal inference under network interference, focusing on identification and estimation under a prespecified treatment assignment scheme \citep{Hudgens2008toward,manski2013identification,liu2016inverse, Aronow2017, basse2018analyzing, Athey2018exact,Leung2020treatment, Forastiere2021,savje2021average, li2022random,Leung2022causal,Hu2022average,ogburn2024causal, belloni2022neighborhood}. 
However, this line of work does not directly address the problem of network experimental design, that is, how to construct a randomization mechanism appropriate for the causal estimand of interest, such as the global treatment effect, spillover effect or direct effect. 
Studying the design problem is important because given a network of connected units, the experimental design determines the joint treatment assignments across the network and directly affects the bias and variance of estimators for causal estimands.

Recent research has investigated experimental design under network interference for different causal estimands. For estimating the global treatment effect, a common strategy is the cluster randomization design, in which the network is partitioned into clusters with minimal interference and treatment is randomly assigned at the cluster level \citep{Ugander2013,eckles2017design,Leung2022rate,Leung2023network,Ugander2023,Liu2022adaptive,Liu2024,viviano2025}. Under such designs, the separation between clusters usually make it difficult to estimate the spillover effect. Moreover, when units within the same cluster share unobserved characteristics that also affect outcomes, cluster-level randomization may fail to balance these characteristics between treatment groups, thereby obscuring the true treatment effect \citep{Shalizi2011,Ugander2023}.
For estimating direct effects and spillover effects, \citet{cai2024} proposed an experimental design based on maximal independent sets, and \citet{Jagadeesan2020design} considered a quasi-coloring design. \citet{viviano2020experimental} and \citet{kandiros2025conflict} proposed designs that can be tailored to different causal estimands. However, the two-wave design in \citet{viviano2020experimental} requires information from a pilot study, which may not be feasible in real applications. \citet{kandiros2025conflict} proposed a novel conflict graph design but the proposed Horvitz-Thompson estimator may have a large variance when exposure probabilities are small.

In this work, we study the ego-cluster randomization design, originally proposed in \cite{saint2019} for large-scale experiments at Linkedin. In this design, the network is partitioned into a collection of ego-clusters, each consisting of a focal unit (the \textit{ego}) and a subset of its immediate neighbors in the network (the \textit{alters}), with treatment randomly assigned at the cluster level; see Figure \ref{fig:egoC} for an example.  
This design enables the simultaneous estimation of global treatment effect and spillover effect. 
In particular, by grouping each ego with its neighbors, the design reduces interference for the egos, facilitating estimation of the global treatment effect. 
Meanwhile, unlike clustering designs that seek to form well-separated larger clusters that reduce interference, the ego-cluster design produces only small clusters, each with an ego and its alters, thereby retaining many cross-cluster connections for egos and alters that are crucial for estimating the spillover effect. 
Despite the successful empirical application of the ego-cluster design \citep{saint2019,su2024}, there is a lack of a statistical framework for studying its theoretical properties and principles for constructing ego-clusters that yield efficient estimation of causal estimands.

To fill this gap, we study the ego-cluster randomization design and analyze its statistical properties for estimating the global treatment effect and spillover effect under a linear outcome model.
Our contributions are threefold. 
First, we establish the consistency and asymptotic normality of regression estimators for the global treatment effect and spillover effect under general dependency-graph conditions.
In contrast to existing ego-cluster designs, which conduct inference only on egos, our analysis uses all sampled units (egos and alters) and is more statistically efficient. To our knowledge, this is the first theoretical analysis of the ego-cluster design. 
Second, we show that the asymptotic variances of the regression estimators are explicit functions of the ego-cluster structure. In particular, the variances depend critically on the fraction of an ego's neighbors that are not assigned to its own ego-cluster, and also on how these external neighbors are distributed across other ego-clusters. Based on this characterization, we develop a two-step greedy algorithm that sequentially selects egos and assigns alters to minimize the asymptotic variance of interest, thereby offering the first principled strategy for constructing ego-clusters. One useful feature of our procedure is that {the experimental design depends only on the network structure and} does not require pre-specifying or tuning the number of clusters. 
Through extensive simulations and two empirical applications, we demonstrate that our proposed experimental design achieves unbiased and efficient estimation for both the global treatment effect and the spillover effect, relative to complete randomization and several existing network experimental designs, and is robust to model misspecification.

Most of the existing work on network experimental design considers a finite-population framework, where the potential outcomes are treated as fixed but unknown quantities and treatment assignment is the only source of randomness in the analysis. This avoids making assumptions on the outcome model or the super-population. However, when the network is only partially observed, as is common in social and online experiments, the estimands from a finite-population framework do not directly generalize to the broader network and may offer limited guidance for implementation. Complementing this line of work, we consider a super-population framework, similar to \citet{Leung2020treatment}, in which the network data are viewed as a sample from an underlying population and can be obtained through, for example, the standard snowball sampling scheme \citep{goodman1961snowball}. We assume a linear outcome model \citep{Basse2018,toulis2013,Parker2017optimal} and estimate the global treatment effect and spillover effect using regression estimators. Regression estimators are commonly used in practice \citep{Leung2020treatment}. 
Nonparametric estimators, such as the Horvitz-Thompson estimator \citep{Ugander2013,Aronow2017,eckles2017design,Leung2020treatment,Leung2022causal,Leung2022rate,Liu2022adaptive,li2022random,Jiang2022statistical,Ugander2023}, are more flexible because they avoid model assumptions, but they may suffer from a small effective sample size after conditioning, leading to high variances.

The remainder of this paper is organized as follows. Section~\ref{sec:frame} introduces our model framework and the ego-cluster design. Section~\ref{sec:theo} develops the theoretical properties of the causal effect estimators. Section~\ref{sec:alg} presents an ego-cluster optimization algorithm. Section~\ref{sec:numerical} reports numerical results, and Section~\ref{sec:real} presents two empirical applications. The paper is concluded with a brief discussion

\section{Model and Experimental Design}\label{sec:frame}
\subsection{Notation, Setup, and Model}\label{sec:setup}
Consider $n$ experimental units indexed by $i\in[n]=\{1,\ldots,n\}$. Each unit receives a binary treatment $T_i\in\{0,1\}$, where $T_i=1$ denotes treatment and $T_i=0$ denotes control. Let $T=(T_1,\ldots,T_n)^\T\in\mathcal{T}=\{0,1\}^n$,
and let $n_1=\sum_{i=1}^{n}T_i$ and $n_0=\sum_{i=1}^{n}(1-T_i)$ denote the numbers of treated and control units, respectively. Suppose we observe a network among the $n$ units, denoted as $G=(V,L)$, where $V=[n]$ is the set of units and $L\subseteq \{(i,j): i,j\in V, i\neq j\}$ is the set of edges between the units. 
Let $A\in \{0,1\}^{n\times n}$ be the symmetric adjacency matrix of $G$. 
For each unit $i$, define its neighboring set $\mathcal{N}_i=\{j: j\in V, A_{ij}=1\}$ and degree $D_i=|\mathcal{N}_i|$. For $d\ge1$, let $\mathcal{N}(i;d)$ denote the set of units whose shortest-path distance from $i$ is at most $d$; in particular, $\mathcal{N}(i;1)=\mathcal{N}_i\cup \{i\}$. 
Let $\rho_i(T,A)=\sum_{j=1}^{n}A_{ij}T_j/D_i$ be the fraction of treated neighbors of unit $i$. When there is no ambiguity, we write $\rho_i$ instead of $\rho_i(T,A)$ for brevity. Let $\rho=(\rho_1,\ldots,\rho_n)^\T$.

For each unit $i$, we observe an outcome $Y_i\in\mathbb{R}$. We assume that $Y_i$ follows
\begin{equation}\label{equ:model}
Y_i=\alpha+\beta T_i+\gamma\rho_i+\epsilon_i,
\end{equation}
where $\epsilon_i$ is the error term satisfying $E(\epsilon_i|A,T)=0$. Similar linear-in-means models have been considered in \citet{toulis2013}, \cite{Cai2015social}, \cite{Parker2017optimal} and \cite{Leung2020treatment}. In model~\eqref{equ:model}, the coefficient $\beta$ captures the contribution of a unit's own treatment to the outcome, while $\gamma$ captures the contribution of its neighbors' treatments as summarized by $\rho_i$. We define $\tau = \beta + \gamma$ as the global treatment parameter, which captures the combined contribution of a unit's own treatment and its neighbors, and define $\gamma$ as the spillover parameter. These two quantities serve as the primary targets of inference for the remainder of the paper. Next, we discuss how $\tau$ and $\gamma$ map to causal estimands.

We adopt the potential outcome framework \citep{Rubin2015} and treat the $n$ units as a random sample from a super population. For each unit $i$, let $Y_i(T)$ denote the potential outcome of unit $i$ under the treatment assignment vector $T\in \mathcal{T}$. The observed outcome is $Y_i=Y_i(T)$ for the realized assignment vector $T$. Model~\eqref{equ:model} can be interpreted as a specification for the potential outcomes \citep{Hu2022average}, \begin{equation}\label{equ:model2}
Y_i(T) = \alpha + \beta T_i + \gamma \rho_i(T,A) + \epsilon_i,
\quad\text{for } T \in \mathcal{T},
\end{equation}
where $\E(\epsilon_i|A,T)=0$.
Under model \eqref{equ:model2}, $Y_i(T)$ depends on $T$ only through the pair $(T_i,\rho_i)$. Equivalently, if two assignment vectors $T,T'\in\mathcal{T}$ satisfy $(T_i,\rho_i) = (T_i',\rho_i')$, then $\E(Y_i(T)) = \E(Y_i(T'))$. With a slight abuse of notation, we may write $Y_i(T)=Y_i(T_i,\rho_i)$. 
This corresponds to \emph{anonymous interference}, in which a unit's potential outcome depends on its neighbors’ treatments only through the fraction of treated neighbors, an assumption that has been widely adopted in the literature 
\citep{Hudgens2008toward,manski2013identification,Leung2020treatment,Forastiere2021,li2022random}.
Correspondingly, we can define the global treatment effect \citep{Hudgens2008toward,kandiros2025conflict}
$$
\tau_{GTE} = \frac{1}{n}\sum_{i=1}^n \E\{Y_i(1,1)-Y_i(0,0)\},
$$
which captures the expected change in potential outcomes when all units are treated versus when none are, 
and the spillover effect \citep{Hudgens2008toward,Forastiere2021}
$$
\tau_{SE} = \frac{1}{n}\sum_{i=1}^n \E\{Y_i(t,1)-Y_i(t,0)\},\qquad t\in\{0,1\},
$$
which captures the expected change in a unit's outcome when all of its neighbors are treated versus when none of its neighbors are, while keeping the unit's own treatment fixed. 
Under model \eqref{equ:model2}, it can be shown that $\tau=\tau_{GTE}$ and $\gamma=\tau_{SE}$.
That is, the parameters $\tau=\beta+\gamma$ and $\gamma$ map to causal estimands as the global treatment effect and the spillover effect regardless of the experiment design.

\subsection{Ego-Cluster Randomization Design}\label{sec:egoc}
\begin{figure}[!t]
\centering
\includegraphics[width=0.55\textwidth]{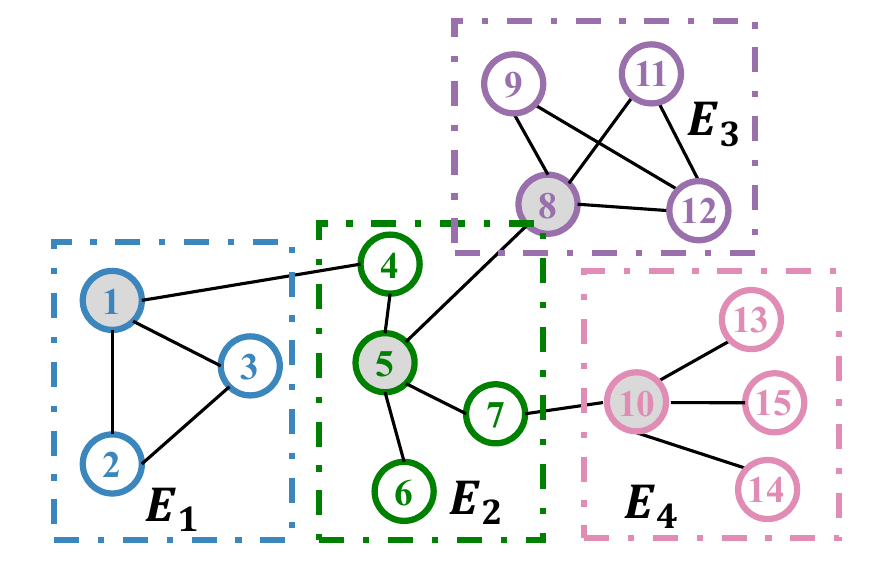}
\caption{Illustration of ego-clusters in a toy network with $15$ units. The network is partitioned into four ego-clusters, $E_1,E_2,E_3,E_4$, represented by four different colors. The ego set consists of units $\{1,5,8,10\}$.}\label{fig:egoC}
\end{figure}

Next, we discuss the ego-cluster randomization design, which divides the $n$ units into a collection of ego-clusters. An ego-cluster consists of a focal unit (the \emph{ego}) and a subset of its neighbors (the \emph{alters}). Correspondingly, within each cluster, the graph distance between any two units is at most two.
We partition the $n$ units into $K_n$ disjoint ego-clusters, 
$$
\mathcal{E}=\left\{\{E_1,\ldots,E_{K_n}\}:\,\,\bigcup_k E_k=[n], \,\,E_k\bigcap E_{k^\prime}=\emptyset \text{ for } k\neq k^\prime\right\}.
$$
The resulting design has $K_n$ egos and $n-K_n$ alters, and $K_n$ may increase with $n$. 
Define $e:[n]\rightarrow[K_n]$ such that $e(i)$ gives the ego-cluster index of unit $i$. Figure~\ref{fig:egoC} gives an example network with four ego-clusters. For example, $e(1)=e(2)=e(3)=1$ and $e(4)=e(5)=e(6)=e(7)=2$. Complete randomization can be seen as a special case of the ego-cluster design with $K_n = n$, that is, each unit forms its own ego-cluster.

\begin{assumption}\label{ass:design}
The sizes of ego-clusters satisfy $\max_{1\leqslant k\leqslant K_n}|E_k|=o(n)$. Conditional on $(A,\mathcal{E})$, the treatment assignments of egos, denoted as $(T_1^\mathcal{E},\ldots T_{K_n}^\mathcal{E})$, are i.i.d. Bernoulli $(1/2)$, and $T_i=T_{e(i)}^\mathcal{E}$ for all $i\in[n]$.
\end{assumption}
Assuming $|E_k|=o(n)$ controls the size of ego-clusters, which is needed for a balanced treatment allocation (see Lemma S1).
Under Assumption~\ref{ass:design}, randomization is conducted at the ego level, and all alters within a cluster receive the same treatment as their ego.

Under model \eqref{equ:model} and Assumption~\ref{ass:design}, the outcome $Y_i$ depends on $T$ through $T_i=T_{e(i)}^\mathcal{E}$ and $\rho_i=\sum_{j=1}^nA_{ij}T_{e(j)}^\mathcal{E}/D_i$. Therefore, given $(A,\mathcal{E})$, the outcome of unit $i$ depends on the treatments of (i) its own ego, and (ii) egos of clusters that contain its neighbors. 
Formally, we write this set as $\mathcal{K}_i=\{k\in[K_n]: \mathcal{N}(i;1)\cap E_k\neq \emptyset\}$. 
In Figure \ref{fig:egoC}, $\mathcal{K}_7=\{2,4\}$, that is, the outcome of unit 7 depends on the treatments of egos in ego-cluster 2 and 4. Under the special case of complete randomization, $\mathcal{K}_i = \mathcal{N}(i;1)$.

Next, we introduce several useful quantities that characterize the network and the ego-cluster structure. 
\textbf{(i) Membership matrix $C$.} Let $C \in \mathbb{R}^{n\times K_n}$ denote a binary cluster membership matrix, where $C_{i,j}=1$ if and only if $e(i)=j$. Since each unit belongs to exactly one ego-cluster, every row of $C$ contains only one nonzero entry. 
\textbf{(ii) Interference matrix $R$.}
Let $R\in \mathbb{R}^{n\times K_n}$ be the interference matrix, where $R_{ik}=|\mathcal{N}_i\cap E_k|/D_i$ gives the proportion of unit $i$'s neighbors that are in ego-cluster $E_k$. 
It follows directly that $R=\text{diag}(D_1,\ldots,D_n)^{-1}AC$, where diag($\cdot$) denotes a diagonal matrix. 
\textbf{(iii) Loss rate $r_i$.}
For a unit $i$, we define its loss rate as $r_i=1-R_{ie(i)}$, which is the fraction of its neighbors that belong to a different ego-cluster than $e(i)$. Let $\bar{r}_n=\sum_{i=1}^{n}r_i/n$ be the average loss rate. Under the special case of complete randomization, we have $C=I_{n\times n}$ and $r_i=1$. In Supplement S1, we provide calculations of $C$, $R$ and $r_i$'s under the example in Figure \ref{fig:egoC}. 
With the above definitions, we can express $\rho_i$, the fraction of treated neighbors for unit $i$, as $\rho_i=\sum_{k=1}^{K_n}R_{ik}T^\mathcal{E}_k$. From this view, $R_{ik}$ quantifies the extent to which $\rho_i$ depends on the treatment assignment of ego-cluster $E_k$.

\subsection{Estimation}
Following the experiment, we estimate the coefficients in model \eqref{equ:model} via regression estimators. 
Let $X=[1_n; T; \rho]$ denote the design matrix, where $1_n$ is an $n$-dimensional vector of ones. 
The estimator can be calculated as
\begin{equation}\label{equ:reg}
(\hat{\alpha},\hat{\beta},\hat{\gamma})^\T=(X^\T X)^{-1}X^\T Y. 
\end{equation}
Correspondingly, the estimators for the global treatment effect and spillover effect are given by $\hat{\tau}=\hat{\beta}+\hat{\gamma}$ and $\hat{\gamma}$, respectively.

To estimate the global treatment effect reliably, it is desirable for the clusters to be well separated, or equivalently, the loss rates $r_i$ to be small, to reduce cross-cluster interference.
On the other hand, well-separated clusters are not suited for estimating the spillover effect. For example, if $r_i=0$ for all units, that is, every neighbor of unit $i$ lies in cluster $e(i)$, we have $\rho_i=T_i$. This is true because $\rho_i=(1-r_i)T^\mathcal{E}_{e(i)}+\sum_{k\neq e(i)}^{K_n}R_{ik}T^\mathcal{E}_k$ and $R_{ik}=0$ for $k\neq e(i)$ in this case.
Estimating both the global treatment effect and spillover effect within a single experiment naturally involves a trade-off: reducing interference improves the estimation of the global treatment effect, whereas too little interference eliminates the variation needed to estimate the spillover parameter $\gamma$.
The ego-cluster design is well-suited for this purpose. 
With appropriately constructed clusters, this design can simultaneously estimate both effects with a single experiment. We discuss how to construct ego-clusters in Section \ref{sec:alg}.

\section{Theoretical Results}\label{sec:theo}

We investigate the theoretical properties of $\hat{\tau}$ and $\hat{\gamma}$ given $(A,\mathcal{E})$.
With network interference, elements of $\{(Y_i,T_i,\rho_i)\}_{i=1}^{n}$ are no longer independent 
across $i\in[n]$. Both $Y_i$ and $\rho_i$ are functions of the treatment assignments of unit $i$ and its neighbors. 
In our analysis, we exclude the trivial case where $r_i=0$ for all $i$.

\subsection{Consistency and Asymptotic Normality}
We first assume error terms $\epsilon_1,\ldots,\epsilon_n$ are i.i.d. as in Assumption~\ref{ass:erroriid}, and then extend our theoretical results to correlated errors in Section~\ref{sec:extension}.
\begin{assumption}\label{ass:erroriid}
Given $A$ and $\mathcal{E}$, the error terms $\epsilon_1,\ldots,\epsilon_n$ are i.i.d. with mean $0$ and variance $\sigma_\epsilon^2$, and are independent of $T$.
\end{assumption}
	
Under Assumption \ref{ass:erroriid}, dependence among elements of $\{(Y_i,T_i,\rho_i)\}_{i=1}^{n}$ is induced by $T$ and $\rho$. Specifically, for unit $i$, $(Y_i,T_i,\rho_i)$ depends on the treatments of egos in $\mathcal{K}_i=\{k\in[K_n]: \mathcal{N}(i;1)\cap E_k\neq \emptyset\}$. We define the dependency graph $G_{\Lambda}=(V,L_{\Lambda})$, where $(i,j)\in L_{\Lambda}$ if and only if $|\mathcal{K}_i \cap \mathcal{K}_j| > 0$, that is, 
$\exists k\in[K_n]$ s.t. $\mathcal{N}(j;1)\cap E_{k}\neq \emptyset\text{ and } \mathcal{N}(i;1)\cap E_{k}\neq \emptyset$. Let $\Lambda\in \mathbb{R}^{n\times n}$ be the adjacency matrix of $G_{\Lambda}=(V,L_{\Lambda})$. Consequently, $(Y_i,T_i, \rho_i)$ and $(Y_j,T_j, \rho_j)$ are dependent if and only if $\Lambda_{ij}=1$, which occurs when (i) units $i,j$ belong to the same ego-cluster; (ii) unit $i$ has neighbors in ego-cluster $e(j)$; (iii) unit $j$ has neighbors in ego-cluster $e(i)$; or (iv) units $i,j$ have neighbors in a common third ego-cluster. By definition, the matrix $\Lambda$ is symmetric and $\Lambda_{ii}=1$. 

We define the dependence set as $N_i=\{j\in V: \Lambda_{ij}=1\}$, such that $(Y_i,T_i, \rho_i)$ is independent of $\{(Y_j,T_j, \rho_j)\}_{j\notin N_{i}}$. For example, in Figure~\ref{fig:egoC}, the dependence set for unit $1$ is $N_{1}=\{2,3,4,5,6,7,8,10\}$. Note that the original graph $G$ specifies neighborhoods $\mathcal{N}_i$'s while the dependency graph $G_{\Lambda}$ specifies neighborhoods $N_i$'s. 
Similar constructions of dependency graphs can be found in \cite{Ugander2013}, \cite{Leung2020treatment}, \cite{Leung2022rate}, \cite{Liu2024} and \cite{viviano2025}. Under the special case of complete randomization, it holds that $K_n=n$ and $\Lambda_{ij}=I(A_{ij}+\max_k A_{ik}A_{kj})$ for $i\neq j$. This experimental design was considered in \cite{Leung2020treatment}.
	
\begin{assumption}\label{ass:omean}
$\sum_{i=1}^{n}|N_i|/n=o(n)$. 
\end{assumption}
	
Assumption~\ref{ass:omean}
controls the amount of interference across ego-clusters, and is satisfied, for example, when each $N_i$
overlaps with a finite number of ego-clusters.
By construction, the maximum graph distance between unit $i$ and any $j\in N_{i}$ is $4$, which occurs when units $i$ and $j$ are connected to two different alters in a common third ego-cluster.
As such, $|N_{i}|$ can be upper bounded by $|\mathcal{N}(i;4)|$, and Assumption~\ref{ass:omean} holds in networks where the average size of $\mathcal{N}(i;4)$ grows at an order of $o(n)$. For example, this holds with high probability for Erd\H{o}s-R\'enyi networks \citep{Erdos1959pmd} with an edge probability of $o(n^{-3/4})$. 

\begin{assumption}\label{ass:net}
{Assume there exists a positive constant $c_0$, such that}
$$
{\max\Big\{\frac{1}{n}\sum_{i=1}^{n}\sum_{k\neq e(i)}^{K_n}R_{ik}^2\,\,,\,\, 
\frac{1}{n}\sum_{i=1}^{n}(r_i-\bar{r}_n)^2\Big\}>c_0.}
$$
\end{assumption}
Assumption~\ref{ass:net} includes two terms: the first term reflects how a unit's lost neighbors are distributed across other ego-clusters and the second term is the variability of loss rates. By $r_i=\sum_{k\neq e(i)}^{K_n}R_{ik}$ and the Cauchy–Schwarz inequality, it holds that $n^{-1}\sum_{i=1}^{n}\sum_{k\neq e(i)}^{K_n}R_{ik}^2\geqslant n^{-1}\sum_{i=1}^{n}r_i^2/(|\mathcal{K}_i|-1)$. Hence, Assumption \ref{ass:net} holds if (i) $r_i=\Theta(1)$ and $|\mathcal{K}_i|=\Theta(1)$ for a non-vanishing fraction of units; or (ii) the variance of $r_i$'s is bounded below by a positive constant. Under complete randomization, each unit has a loss rate of 1 and $R_{ik}=1/D_i$ for $k\in \mathcal{N}_i$.
Consequently, the first term reduces to  $\sum_{i=1}^n (1/D_i)/n$, while the second term equals $0$.
Hence, a sufficient condition for Assumption \ref{ass:net} under complete randomization is that a non-vanishing fraction of units have a bounded degree (by a constant). 
Finally, the term in Assumption~\ref{ass:net} admits a natural upper bound that is $\max\{ \sum_{i=1}^{n}r_i^2/n,\sum_{i=1}^{n}(r_i-\bar{r}_n)^2/n\}$, which is uniformly upper bounded as $0 \leqslant r_i \leqslant 1$ for all $i\in[n]$.

\begin{theorem}\label{thm:consistency}
Suppose Assumptions~\ref{ass:design}-~\ref{ass:net} hold, then $\hat{\tau}\stackrel{P}{\rightarrow}\tau$ and $\hat{\gamma}\stackrel{P}{\rightarrow}\gamma$.
\end{theorem}
	
Theorem~\ref{thm:consistency} establishes the estimation consistency of $\hat{\tau}$ and $\hat{\gamma}$. To derive their asymptotic normality, we employ Stein's method \citep{ross2011,Leung2020treatment}, which requires further assumptions. Let $(\Lambda^3)_{ij}$ denote the $(i,j)$-th entry of matrix $\Lambda^3$.
    
\begin{assumption}\label{ass:omax}
$\sum_{i=1}^{n}|N_i|^2/n=o(\sqrt{n})$, $\sum_{i=1}^{n}|N_i|^3/n=o(n)$, $\sum_{i,j=1}^{n}(\Lambda^3)_{ij}/n=o(n)$.
\end{assumption}

Assumption~\ref{ass:omax} imposes higher-order moment conditions on the dependence set compared to Assumption~\ref{ass:omean}. It is easy to verify that $\sum_{i=1}^{n}|N_i|^2/n\leqslant (\max_i |N_i|)^2$, $\sum_{i=1}^{n}|N_i|^3/n\leqslant (\max_i |N_i|)^3$ and $\sum_{i,j=1}^{n}(\Lambda^3)_{ij}/n\leqslant (\max_i |N_i|)^3$. Therefore, a sufficient condition for Assumption~\ref{ass:omax} is that the maximum degree of the dependency graph satisfies $\max_i |N_i|=o(n^{1/4})$.

\begin{theorem}\label{thm:normal}
Suppose Assumptions~\ref{ass:design},~\ref{ass:erroriid},~\ref{ass:net} and~\ref{ass:omax} hold. {Write $b_n=\sum_{i=1}^{n}\sum_{k\neq e(i)}^{K_n}R_{ik}^2/n+\sum_{i=1}^{n}(r_i-\bar{r}_n)^2/n$.} If $\max_i\E(\epsilon_i^4|A,\mathcal{E})<\infty$
then
$$\sigma_{\tau,n}^{-1}\sqrt{n}(\hat{\tau}-\tau)\stackrel{d}{\rightarrow}\mathcal{N}(0,1),\quad\sigma_{\gamma,n}^{-1}\sqrt{n}(\hat{\gamma}-\gamma)\stackrel{d}{\rightarrow}\mathcal{N}(0,1),$$
where $\sigma_{\tau,n}^{2}=4\sigma_\epsilon^2(\bar{r}_n^2/b_n+1)$, $\sigma_{\gamma,n}^{2}=4\sigma_\epsilon^2/b_n$.
\end{theorem}

Theorem~\ref{thm:normal} establishes the asymptotic normality of $\hat{\tau}$ and $\hat{\gamma}$, whose variances depend on the underlying ego-cluster structure.
As the average loss rate $\bar r_n$ reflects the overall level of interference, it is seen that low-interference ego-clusters are preferable for estimating the global treatment effect.
One interesting observation is that, for any fixed average loss rate $\bar r_n>0$, both variances decrease with $b_n$.
This may appear counter-intuitive, as we discussed earlier that there is a trade-off between estimating the global treatment effect and the spillover effect within a single experiment. However, the average loss rate $\bar r_n$ and $b_n$ cannot be optimized in the ego-cluster design separately. Recall that $b_n\leqslant \sum_{i=1}^{n}r_i^2/n+\sum_{i=1}^{n}(r_i-\bar{r}_n)^2/n$. Since $0 \le r_i \le 1$ implies $r_i^2 \le r_i$, we have $\frac{1}{n}\sum_{i=1}^n r_i^2 \le \frac{1}{n}\sum_{i=1}^n r_i = \bar r_n$. Given $\frac{1}{n}\sum_{i=1}^n (r_i - \bar r_n)^2= \frac{1}{n}\sum_{i=1}^n r_i^2 - \bar r_n^2$, it holds that $b_n\le2\bar r_n - \bar r_n^2= \bar r_n(2 - \bar r_n)\le2\bar r_n$. Hence, if the average loss rate $\bar r_n$ is small, then $b_n$ must also be small. In this case, $\sigma_{\gamma,n}^{2}$ becomes large whereas $\sigma_{\tau,n}^{2}$ may be large or small, depending on the ratio of $\bar r_n^2/b_n$.

Theorem~\ref{thm:normal} enables improving the statistical efficiency of $\hat\tau$ and $\hat\gamma$ by appropriately designing the ego-clustering algorithm. The two variances suggest different optimization objectives for different causal estimands. 
If the focus is on estimating the global treatment effect $\tau$, one can aim to minimize ${\bar r_n^2}/{b_n}$; if the focus is on the spillover effect $\gamma$, one may instead minimize $1 / b_n$, thereby placing emphasis on maximizing $b_n$. When both $\tau$ and $\gamma$ are of interest, a weighted objective such as $\lambda \bar{r}_n^2 / b_n + (1 - \lambda) / b_n$, for some $0 < \lambda < 1$, can be employed to balance the trade-off between the two estimands.

\begin{remark}
{We note that the ego-clustering algorithm in Section \ref{sec:alg} depends only on the network structure $A$ and not on the observed outcomes $Y$ or treatment $T$. Consequently, the partition $\mathcal{E}$ is fixed before any outcomes are realized, and the asymptotic results in Theorems \ref{thm:consistency} and \ref{thm:normal} apply directly without requiring sample splitting between design and inference.}
\end{remark}

\subsection{Confidence Intervals and Hypothesis Testing}\label{sec:inf}
{To enable valid inference, we provide a consistent estimator for $\sigma_{\epsilon}^2$. Let $(\hat{\epsilon}_1,\ldots,\hat{\epsilon}_n)$ be the residuals from the regression under Model (\ref{equ:model}), that is $\hat{\epsilon}_i=Y_i-\hat{\alpha}-\hat{\beta}T_i-\hat{\gamma}\rho_i$. We estimate $\sigma_{\epsilon}^2$ by $\hat{\sigma}_{\epsilon}^2=\sum_{i=1}^{n}\hat{\epsilon}_i^2/(n-3)$ and establish the consistency of $\hat{\sigma}_{\epsilon}^2$ in the following theorem.}

\begin{theorem}\label{thm:varest}
{Suppose assumptions in Theorem~\ref{thm:normal} hold. Then
$\hat{\sigma}_{\epsilon}^2\stackrel{P}{\rightarrow}\sigma_{\epsilon}^2$.}
\end{theorem}

{Under Theorems \ref{thm:normal}-\ref{thm:varest}, $100(1-\alpha)$\% confidence intervals of $\hat{\tau}$ and $\hat{\gamma}$ can be constructed as
    $$\hat{\tau}\pm z_{\alpha/2}\frac{2\hat{\sigma}_\epsilon\sqrt{\bar{r}_n^2/b_n+1}}{\sqrt{n}},\quad \hat{\gamma}\pm z_{\alpha/2}\frac{2\hat{\sigma}_\epsilon\sqrt{1/b_n}}{\sqrt{n}},$$
respectively, where $z_{\alpha/2}$ denotes the upper $\alpha/2$ quantile of $N(0,1)$.
To test
$$H_{0,\tau}: \tau=0\text{ v.s. } H_{1,\tau}: \tau\neq0 \quad\text{or}\quad H_{0,\gamma}: \gamma=0\text{ v.s. } H_{1,\gamma}: \gamma\neq0,$$
we consider test statistics
$$T_{\tau}=\frac{\sqrt{n}\hat{\tau}}{2\hat{\sigma}_{\epsilon}\sqrt{\bar{r}_n^2/b_n+1}},\quad\text{or}\quad T_{\gamma}=\frac{\sqrt{n}\hat{\gamma}}{2\hat{\sigma}_{\epsilon}\sqrt{1/b_n}},$$
respectively. 
By Theorems~\ref{thm:normal} and \ref{thm:varest}, both $T_{\tau}$ and $T_{\gamma}$ are asymptotically $N(0,1)$ under the null. At significance level $\alpha$, we reject $H_{0,\tau}$ if $|T_{\tau}|>z_{\alpha/2}$ and reject $H_{0,\gamma}$ if $|T_{\gamma}|>z_{\alpha/2}$. The corresponding two-sided p-values are given by $p_{\tau}=2\{1-\Phi(|T_{\tau}|)\}$ and $p_{\gamma}=2\{1-\Phi(|T_{\gamma}|)\}$, respectively, where $\Phi(\cdot)$ is the cumulative distribution function of $N(0,1)$.}

\subsection{Extension to Correlated Errors}\label{sec:extension}
In this section, we extend the theoretical results to settings with correlated random errors. Denote $\nu_{i}^2=\Var(\epsilon_i)$ and $\nu_{ij}=\Cov(\epsilon_i,\epsilon_j)$.
	
\begin{assumption}\label{ass:error}
{For any} given $A$ and $\mathcal{E}$, consider the following assumptions:\\
(a) $\epsilon_i\perp \epsilon_j$ if $A_{ij}=\max_k A_{ik}A_{kj}=0$ for $i\neq j$;\\
(b) $\epsilon\perp T$, $\E(\epsilon)=0$, $\max_{1\leqslant i\leqslant n}\nu_{i}^2<\infty$;\\
(c1) $\sum_{i=1}^{n}\sum_{j\in \mathcal{N}(i;2)/\{i\}}\nu_{ij}/n=o(n)$;\\
(c2) $\sum_{i=1}^{n}\sum_{j\in \mathcal{N}(i;2)/\{i\}}\nu_{ij}/n=O(1)$.
\end{assumption}
Assumption~\ref{ass:error}(a) specifies the correlation structure of the error terms, allowing $\epsilon_i$ and $\epsilon_j$ to be correlated when they are directly connected ($A_{ij}=1$) or share a common neighbor ($\max_k A_{ik}A_{kj}=1$).
If the distance between $i$ and $j$ exceeds two, their errors are assumed independent. This local dependence structure is commonly used in the literature \citep{Leung2020treatment, Liu2024, ogburn2024causal}. 
Assumption~\ref{ass:error}(b) requires that, given the network and ego-clusters, the error terms are independent of treatment assignment $T$, and have mean zero and finite variances. Assumption~\ref{ass:error}(c1) is introduced to derive the consistency of causal estimators, while establishing asymptotic normality requires the tighter order imposed by Assumption~\ref{ass:error}(c2). Assumptions~\ref{ass:error}(c1)-(c2) are similar to the conditions used in \cite{Leung2020treatment}. Under homogeneous covariance where $\nu_{ij}=\nu$, Assumptions~\ref{ass:error}(c1)-(c2) simplify to order conditions on the average number of neighbors within a graph distance of two, as $\sum_{i=1}^{n}|\mathcal{N}(i;2)/\{i\}|/n=\sum_{i=1}^{n}|\mathcal{N}(i;2)|/n-1$.

Compared with Assumption~\ref{ass:erroriid}, Assumption~\ref{ass:error} introduces an additional source of dependence among the observed data $\{(Y_i,T_i,\rho_i)\}_{i=1}^n$. 
However, the dependency graph $\Lambda$ defined in Section~\ref{sec:theo} remain unchanged, as $\Lambda_{ij}=1$ whenever $A_{ij}=1$ or $\max_k A_{ik}A_{kj}=1$.

\begin{theorem}\label{thm:consistency2}
Suppose Assumptions~\ref{ass:design},~\ref{ass:omean},~\ref{ass:net},~\ref{ass:error}(a)-(c1) hold, then $\hat{\tau}\stackrel{P}{\rightarrow}\tau$ and $\hat{\gamma}\stackrel{P}{\rightarrow}\gamma$.
\end{theorem}

\begin{theorem}\label{thm:normal2}
Suppose Assumptions~\ref{ass:design},~\ref{ass:net},~\ref{ass:omax} and~\ref{ass:error} hold. If $\max_i\E(\epsilon_i^4|A,\mathcal{E})<\infty$, then 
$$\sigma_{\tau,n}^{-1}\sqrt{n}(\hat{\tau}-\tau)\stackrel{d}{\rightarrow}\mathcal{N}(0,1),\quad\sigma_{\gamma,n}^{-1}\sqrt{n}(\hat{\gamma}-\gamma)\stackrel{d}{\rightarrow}\mathcal{N}(0,1),$$
where the explicit forms of $\sigma_{\tau,n}^2$ and $\sigma_{\gamma,n}^2$ are provided in Supplement 2.3.
\end{theorem}

Theorems~\ref{thm:consistency2} and~\ref{thm:normal2} establish the consistency and asymptotic normality of $\hat{\tau}$ and $\hat{\gamma}$ when the error terms are correlated. As shown in Supplement 2.3, the variances have complex expressions of the form $\sigma_{\tau,n}^{2}=\frac{1}{n}\sum_{i=1}^{n}\E(W_i^2)\nu_{i}^2+\frac{1}{n}\sum_{i=1}^{n}\sum_{j\in \mathcal{N}(i;2)/\{i\}}\E(W_iW_j)\nu_{ij}$ {and $\sigma_{\gamma,n}^{2}=\frac{1}{n}\sum_{i=1}^{n}\E(W_i^{\prime2})\nu_{i}^2+\frac{1}{n}\sum_{i=1}^{n}\sum_{j\in \mathcal{N}(i;2)/\{i\}}\E(W_i^\prime W_j^\prime)\nu_{ij}$}, where $(W_1,\ldots,W_n)$ and $(W_1^\prime,\ldots,W_n^\prime)$ are random variables determined by $A$, $\mathcal{E}$ and $T$. When the errors are i.i.d. as specified in Assumption~\ref{ass:erroriid}, that is $\nu_{ij}=0$ and $\nu_{i}^2=\sigma_\epsilon^2$, the cross-covariance terms vanish, and Theorem~\ref{thm:normal2} reduces to the i.i.d. case in Theorem~\ref{thm:normal}. {In the presence of correlated errors, variance estimation can be performed using the generalized robust standard errors as proposed by \cite{Leung2020treatment}. Further details are discussed in Supplement S2.4.}

\section{Ego-Clustering Algorithm}\label{sec:alg}
In this section, we focus on the global treatment effect $\tau$ and adopt $\bar{r}_n^2 / b_n$ as the minimization target in our algorithm, which guides the construction of ego-clusters with a small value of $\bar{r}_n$ and a large value of $b_n$. Our numerical studies demonstrate that our design also performs well in estimating the spillover effect. Other targets such as $1/b_n$ or $\lambda \bar{r}_n^2 / b_n + (1 - \lambda) / b_n$, for $0 < \lambda < 1$ can be adopted as well.

While a joint optimization of ego selection and alter assignment is conceptually possible, it leads to a combinatorial problem with an exponentially large search space and is computationally infeasible for networks of practical sizes. This motivates a greedy two-step procedure that first selects egos and then reassigns alters, with the goal to minimize $\bar{r}_n^2/b_n$.
Different from existing methods \citep{saint2019,su2024,deng2024}, which typically select egos at random and assign alters based on edge weights, our procedure is guided directly by the estimator variance $\bar{r}_n^2 / b_n$ derived in our theoretical results. The proposed ego-clustering algorithm also differs conceptually from graph-theoretic clustering methods such as 3-net \citep{Ugander2013}. While 3-net constructs clusters based on metric separation in the network, our approach is explicitly guided by a variance objective derived from the causal estimand.

\begin{figure}[!t]
\centering
\includegraphics[width=0.85\textwidth]{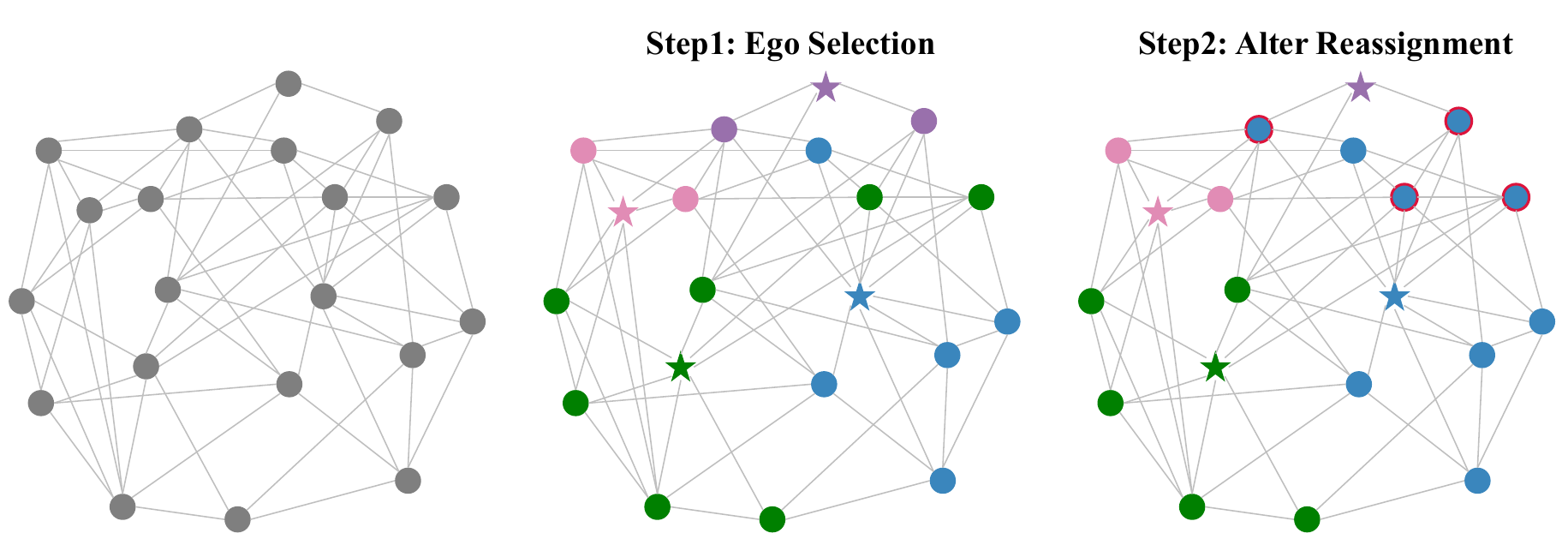}
\caption{Illustration of the two-step ego-clustering algorithm. (Left) A small-world network. (Middle) Step 1: Ego units (stars) are selected to form temporary ego-clusters. (Right) Step 2: Four alters (with red border) are reassigned.}\label{fig:algorithm}
\end{figure}

\subsection{Ego Selection}\label{sec:alg1}
We consider a backward-selection algorithm for choosing egos, starting from the case where every unit is an ego. Correspondingly, the initial configuration has $K_n=n$, $C=I_n$, and $R=\text{diag}(D_1,\ldots,D_n)^{-1}A$. The algorithm then proceeds by iteratively forming ego-clusters in a way that decreases the objective value.

At each iteration, we randomly select a candidate ego from the units not yet assigned to any ego-cluster and form a candidate ego-cluster with its currently unassigned neighbors; already-assigned alters are not reassigned. If this candidate ego-cluster decreases the objective, it is accepted; otherwise, it is discarded and another ego is drawn. The process repeats until no further reduction is possible, with any unselected unit retaining its initial singleton ego-cluster. Algorithm~\ref{algorithm1} summarizes the procedure, and Figure~\ref{fig:algorithm} gives an illustration.

The number of columns in $C$ reflects the current number of ego-clusters. As the algorithm proceeds and new ego-clusters are formed, the number of columns decreases from its initial value of $n$ to the final value $K_n$. Consequently, the algorithm automatically determines the appropriate number of clusters $K_n$ in a data-driven manner. This eliminates the need to specify the number of clusters prior to the experiment.
In some cases, certain units may be predetermined to serve as egos. In such cases, we can set these units as initial egos and then choose other egos following our algorithm.

Since $C$ and $R$ are sufficient for evaluating the objective, we update only their affected entries at each step. Rows and columns are initially indexed by $[n]$; when ego $m$ is grouped with its unassigned alters $\text{alters}_m$, the columns labeled by $\text{alters}_m$ are summed row-wise into column $m$ and then removed. {Each selection step requires $O(ND_m)$ operations.}

	\begin{algorithm}[!t]
		\caption{Ego Selection Algorithm}\label{algorithm1}
		\KwIn{Adjacency matrix $A$, membership matrix $C = I_n$, interference matrix $R$, initial objective value $obj$.}
		\KwOut{Updated membership matrix $C$.}
		
		Initialize \textit{ego\_set} $\leftarrow \emptyset$, \textit{candidates} $\leftarrow [n]$,
		\textit{stop} $\leftarrow \textit{False}$\;
		
		\Repeat{\textit{stop} = \textit{True}}{
			\textit{node\_pool} $\leftarrow$ \textit{candidates}, \textit{selected} $\leftarrow \textit{False}$\;
			
			\While{\textit{selected} = \textit{False}}{
				\If{\textit{node\_pool} is $\emptyset$}{
					\textit{stop} $\leftarrow \textit{True}$; \textbf{break}\;
				}
				
				Randomly select an ego from \textit{node\_pool} and group it with its unassigned neighbors to form a candidate ego-cluster\;
				Compute current objective value $obj_{new}$\;
				
				\eIf{$obj_{new} < obj$}{
					Accept this candidate ego-cluster, \textit{selected} $\leftarrow \textit{True}$\;
					Update $C$, $R$, $obj=obj_{new}$, and delete this ego-cluster from \textit{candidates}\;
					
				}{
					Discard the candidate ego-cluster and delete this ego from \textit{node\_pool}\;
				}
			}
		}
	\end{algorithm}

\subsection{Alter Reassignment}\label{sec:alg2}
Given the ego-clusters from the previous step, we keep the egos fixed and reassign the alters to further reduce the objective. Consider an alter unit $m$ that is connected to multiple egos, including $k_1$ and $k_2$, and is currently assigned to ego-cluster $E_{k_1}$. To evaluate whether $m$ should be reassigned to $E_{k_2}$, we examine the resulting change in the objective value. As discussed earlier, it is sufficient to analyze changes in the membership matrix $C$ and interference matrix $R$. Let $C^\ast$ and $R^\ast$ denote the updated matrices after reassignment. In the membership matrix $C$, this reassignment corresponds to setting $C_{m,k_1}^\ast=0$ and $C_{m,k_2}^\ast=1$. For the interference matrix $R$, the updates are given by $$R_{hk_1}^\ast=R_{hk_1}-\frac{1}{D_h},\quad R_{hk_2}^\ast=R_{hk_2}+\frac{1}{D_h}, \quad \text{for }h\in \mathcal{N}_m.$$
These adjustments to $R$ lead to corresponding updates in the loss rates: $r_{h}^\ast=r_{h}+\frac{1}{D_h}$ for $h\in E_{k_1}\cap\mathcal{N}_m$; $r_{h}^\ast=r_{h}-\frac{1}{D_h}$ for $h\in E_{k_2}\cap\mathcal{N}_m$; and $r_{m}^\ast=r_{m}+\frac{|\mathcal{N}_m\cap E_{k_1}|}{D_m}-\frac{|\mathcal{N}_m\cap E_{k_2}|}{D_m}$.

The objective depends on two key quantities: $\bar{r}_n$ and $b_n$. Their values after reassignment are directly given by:
$$
\begin{aligned}
\bar{r}^\ast_n=&\bar{r}_n+\frac{1}{n}\left(\sum_{h\in E_{k_1}\cap\mathcal{N}_m}\frac{1}{D_h}-\sum_{h\in E_{k_2}\cap\mathcal{N}_m}\frac{1}{D_h}+\frac{|\mathcal{N}_m\cap E_{k_1}|-|\mathcal{N}_m\cap E_{k_2}|}{D_m}\right),\\
b_n^\ast=&b_n+\frac{2}{n}\sum_{h\in \mathcal{N}_m}\left(\frac{1}{D_h^2}-\frac{R_{hk_1}-R_{hk_2}}{D_h}\right)+(1-\bar{r}_n)^2-(1-\bar{r}_n^\ast)^2,
\end{aligned}
$$
which is fast to evaluate. Using these updates, we accept the reassignment if the resulting objective value decreases. 
{The computational complexity of the alter reassignment step for node $m$ is $O(D_m)$.} This check is repeated for all alters connected to multiple egos. Algorithm~\ref{algorithm2} summarizes the alter reassignment step, and Figure~\ref{fig:algorithm} gives an illustration.

	\begin{algorithm}[!t]
		\caption{Alter reassignment}\label{algorithm2}
		\KwIn{Adjacency matrix $A$, membership matrix $C$, interference matrix $R$, objective value $obj$}
		\KwOut{Updated membership matrix $C$}
		\textit{Alter\_considered}$\leftarrow\{\text{alters connected to two or more egos}\}$\;
        Randomly permute \textit{Alter\_considered}\;
		\Repeat{no alter is reassigned}{
			\ForEach{alter $m\in$ \textit{Alter\_considered}}{
				Find connected egos' indices \textit{$Egos_m$} and the current ego-cluster index $e(m)$\;
				\ForEach{$j\in Egos_m$}{
					Calculate updated objective value $obj_{new}$ if $m$ is reassigned to $E_j$\;
					\If{$obj_{new}<obj$}{
						Reassign alter $m$ from $E_{e(m)}$ to $E_j$\;
						update $C$, $R$, $e(m)=j$ and $obj=obj_{new}$\;
					}
				}
			}
		}
	\end{algorithm}

\section{Numerical Studies}\label{sec:numerical}
In this section, we examine the finite sample performance of our proposed method, referred to as \texttt{EgoCR}, in estimating the global treatment effect and the spillover effect under several network settings. We begin with two random network models: Erd\H{o}s--R\'enyi model \citep{Erdos1959pmd}, denoted as ER($n, p$), where $p$ is the connecting probability between any two units; and Barab\'asi–Albert network \citep{BA1999emergence}, denoted as BA($n,m$), where $m$ represents the number of edges that each new unit forms when entering the network. The outcome model is $Y=\alpha+\beta T+\gamma \rho+\epsilon$. The parameters are set as $\alpha=2$, $\beta=2.5$, $\gamma=5$, with the error terms $\epsilon_i$'s independently drawn from a standard normal distribution. To better capture features commonly observed in real social networks, we also consider a network composed of four communities, each generated using the Small-World model \citep{watts1998collective}. To introduce cross-community connectivity, sparse inter-community edges are added randomly such that the average probability of forming a within-community edge is eight times that of forming a cross-community edge. 
We refer to this design as the community network, CN($n$). Since individuals within the same community tend to share similar characteristics, we further introduce an unobserved confounder $Z=(Z_1,\ldots,Z_n)$, which varies across different communities and incorporate it into the outcome model as $Y=\alpha+\beta T+\gamma \rho+\eta Z+\epsilon$. We set $\eta=0.8$ in our experiments. 
	
For comparison, we implement several alternative designs: complete randomization (\texttt{CR}); the Conflict Graph Design of \citet{kandiros2025conflict} (\texttt{CGD}); the causal clustering method of \citet{viviano2025} (\texttt{CausalC}); the $3$-net clustering of \citet{Ugander2013} (\texttt{$3$-net}); the ego-cluster design of \citet{saint2019} (\texttt{LinkedIn}); Louvain clustering \citep{Blondel2008} using the default settings in the igraph package (\texttt{Louvain}); and the 1-hop-max clustering of \citet{Ugander2023} (\texttt{RGCR1hm}). For the ER network, we set the connection probability to $p=15/n$; for the BA network, we set $m=6$; and for the CN network, we target an average degree of approximately $11$. For each scenario, we generate 2000 networks and conduct 2000 experimental replications. 

\begin{figure}[!t]
\centering
\includegraphics[width=0.8\textwidth]{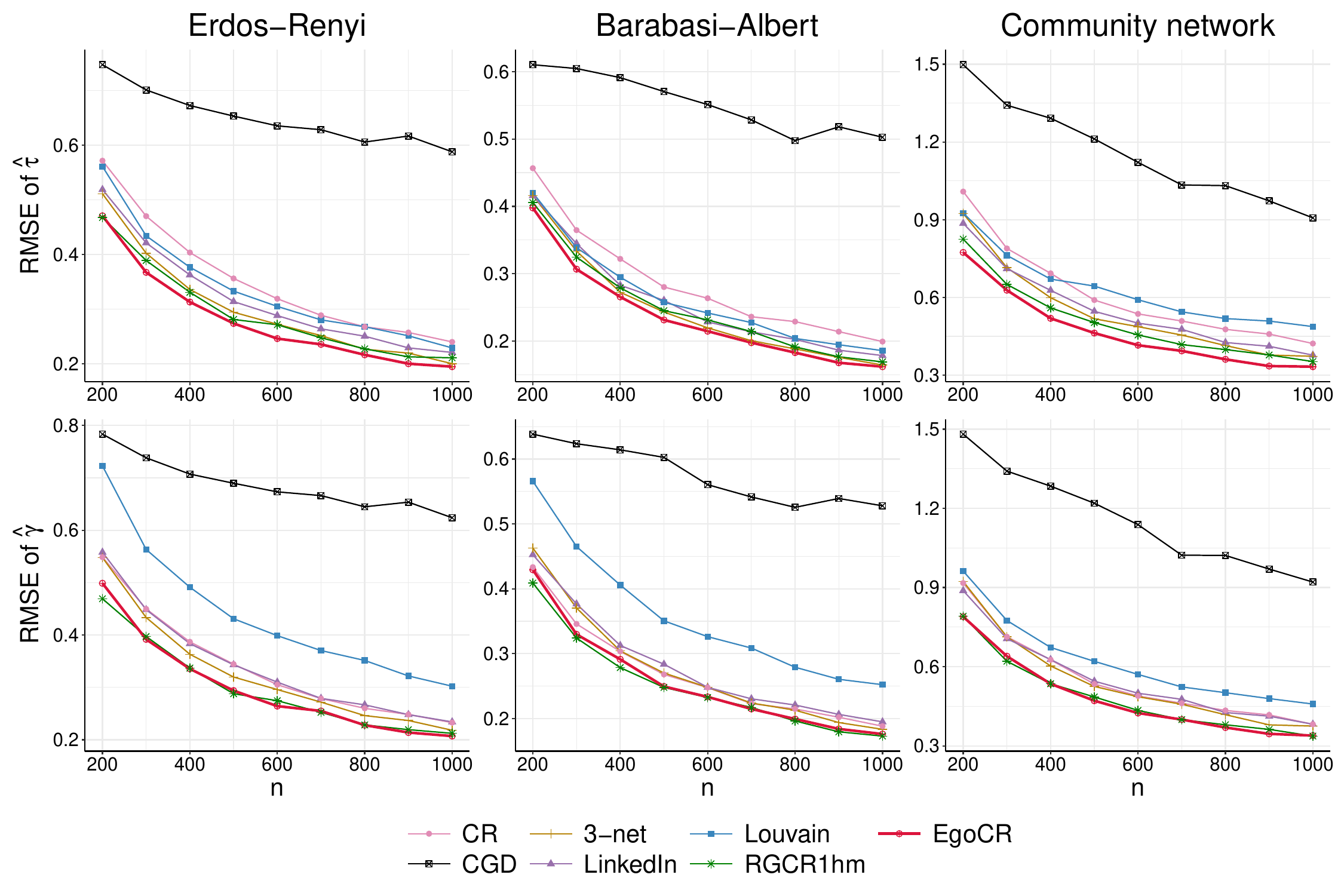}
\caption{RMSEs of $\hat{\tau}$ (top) and $\hat{\gamma}$ (bottom) under different sample sizes and clustering methods over $2000$ replications for each network setting. \label{fig:plot_rmse}}
\end{figure}

Figure~\ref{fig:plot_rmse} reports the root mean squared errors (RMSEs) of $\hat{\tau}$ and $\hat{\gamma}$ for sample sizes ranging from $n=200$ to $1000$; comparison of biases {with standard errors} are included in Supplement S3.1. Table S1 {in Supplement S3.1} provides additional comparisons under alternative network configurations for $n=500$ and $n=1000$ (with average connection probabilities around $0.05$). To facilitate a consistent comparison with our approach, we report regression-based estimators for \texttt{CGD}, \texttt{CausalC}, and \texttt{LinkedIn} in the main text, while the results for their original estimators are included in the Supplement. For \texttt{CausalC}, we only include the results for $n=500$ in Table S1 because of its high computational demand.

As shown in Table S1 and Figure S1 (Supplement S3.1), \texttt{EgoCR} has very small biases for both $\hat{\tau}$ and $\hat{\gamma}$ across all settings considered. For estimating the global treatment effect, Figure~\ref{fig:plot_rmse} shows that \texttt{EgoCR} achieves the lowest RMSEs as the sample size increases. In Figure~\ref{fig:plot_rmse} (top), the cluster-based approaches generally outperform \texttt{CR}, as clustering helps to reduce interference. \texttt{CGD} performs less favorably in these settings, likely because the maximum eigenvalue of the conflict graph ($A^2$ in \citet{kandiros2025conflict}) is large for these networks, which in turn leads to a very limited number of treated assignments. 
\texttt{LinkedIn} design yields higher RMSEs than \texttt{EgoCR}, because it forms ego-clusters almost randomly, whereas our method targets to minimize estimator variance. Moreover, \texttt{CausalC} is not designed for regression estimators, and as a result, it yields larger RMSEs compared with \texttt{EgoCR} as shown in Table S1.

Regarding the spillover effect, Figure~\ref{fig:plot_rmse} shows that \texttt{EgoCR} achieves smaller RMSEs than alternative methods except \texttt{RGCR1hm}. \texttt{RGCR1hm}, proposed by \cite{Ugander2023}, demonstrates comparable performance of $\hat{\gamma}$ to \texttt{EgoCR} in Figure~\ref{fig:plot_rmse}. However, in Table S1, which reports results for denser network settings, \texttt{EgoCR} gives a higher accuracy than \texttt{RGCR1hm}.
\texttt{Louvain} produces larger RMSEs of $\hat{\gamma}$ than \texttt{CR}, suggesting that clusters constructed by tightly connected communities are not suited for estimating spillover effects.

In the community network settings, we introduce an underlying confounder $Z$ that varies across communities.
Since community detection algorithms tend to form clusters within these communities, randomization at the cluster level may induce an imbalance in the confounder between treatment arms, thereby increasing the variability of estimators. 
In contrast, our proposed method forms a large number of ego-clusters based on local neighborhoods rather than the community structure. This reduces the influence of this confounder and results in smaller RMSEs than most alternative methods.

\begin{figure}[!t]
		\centering
		\includegraphics[width=0.8\textwidth]{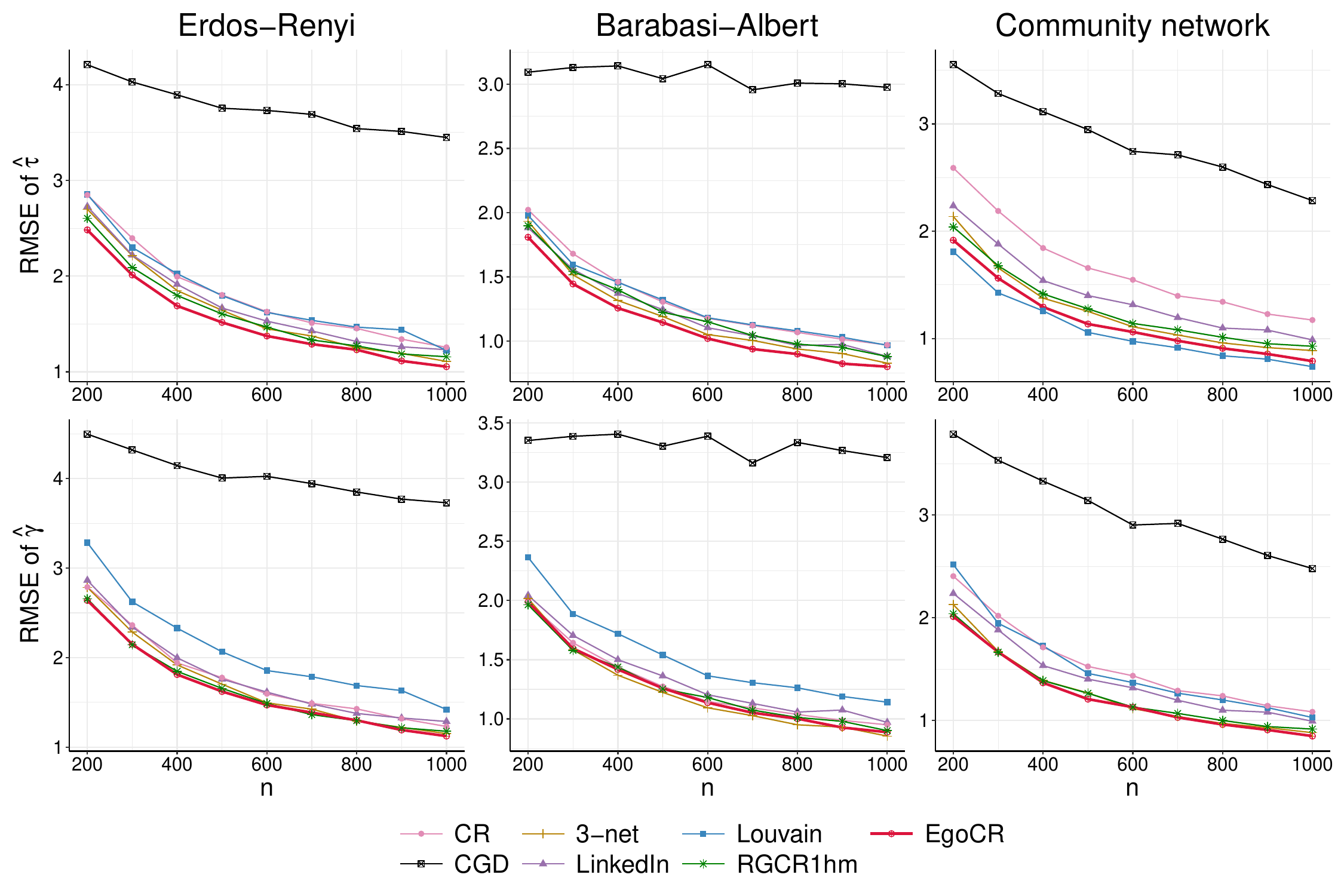}
		\caption{RMSEs of $\hat{\tau}$ (top) and $\hat{\gamma}$ (bottom) under different sample sizes and clustering methods over $2000$ replications for each network setting under correlated errors. \label{fig:plot_rmse_corerror}}
\end{figure}

{Under the same settings as in Figure~\ref{fig:plot_rmse}, we replace the i.i.d. error terms with correlated errors, where $\epsilon=A\epsilon^\prime$ and $\epsilon^\prime$ follows an i.i.d. standard normal distribution. The resulting RMSEs of $\hat{\tau}$ and $\hat{\gamma}$, reported in Figure~\ref{fig:plot_rmse_corerror}, indicate that our proposed method, \texttt{EgoCR}, continues to provide satisfactory estimation of both $\tau$ and $\gamma$. We have also considered other misspecified settings where the outcome does not follow Model (\ref{equ:model}). These results are included in Supplement S3.2 and lead to similar conclusions.} 

{We then evaluate the performance of the inferential methods proposed in Section~\ref{sec:inf}. We consider the community network setting in Figure~\ref{fig:plot_rmse} and the Barab\'asi–Albert network with parameter $m=5$, for sample sizes $n=500$ and $1000$. The nominal significance level is $\alpha=5$\%. The rejection probabilities for testing $H_{0,\tau}$ and $H_{0,\gamma}$ are reported in Table~\ref{tab:inference}. When $\tau=0$ ($\gamma=0$), the simulated Type I errors under \texttt{EgoCR} approach the nominal level of 5\%, whereas those under \texttt{Louvain} and \texttt{3-net} are inflated. Under the alternative hypotheses, \texttt{EgoCR} demonstrates favorable rejection probabilities compared to other methods.

\begin{table}[!t]
		\centering
		\caption{Rejection probabilities (\%) for testing $H_{0,\tau}$ and $H_{0,\gamma}$ under different sample sizes and clustering methods over $2000$ replications for each network setting. \label{tab:inference}}
		\scalebox{0.7}{
				\begin{tabular}{llrrrrrrrrrrrr}
					\toprule
					\toprule
                    & & \multicolumn{6}{c}{Barab\'asi–Albert} & \multicolumn{6}{c}{Community Network} \\
                    \cmidrule(lr){3-8}\cmidrule(lr){9-14}
					 Method & $n$& $\tau=0$ & $\tau=0.3$ & $\tau=0.6$ & $\gamma=0$ & $\gamma=0.3$ & $\gamma=0.6$ & $\tau=0$ & $\tau=0.3$ & $\tau=0.6$ & $\gamma=0$ & $\gamma=0.3$ & $\gamma=0.6$\\
					\midrule
                    CR&500 & 6.30 & 20.90 & 63.40 & 6.30 & 23.40 & 68.00& 6.30 & 17.20 & 50.00 & 6.40 & 18.75 & 52.50\\
                    &1000 & 5.70 & 36.95 & 89.75 & 5.10 & 41.10 & 93.70& 5.30 & 29.15 & 79.45 & 5.30 & 31.70 & 84.25\\
                    \midrule
                     $3$-net &500 & 9.25 & 38.50 & 86.60 & 8.40 & 31.40 & 78.10 & 9.55 & 42.00 & 89.60 & 6.80 & 30.55 & 74.05\\
                    &1000 & 7.90 & 59.20 & 98.10 & 7.95 & 48.30 & 94.15& 7.20 & 62.70 & 99.00 & 6.35 & 47.40 & 95.40\\
                    \midrule
                    LinkedIn &500 & 6.20 & 27.95 & 76.35 & 5.95 & 23.35 & 67.90& 5.65 & 26.55 & 73.85 & 5.40 & 21.60 & 63.05\\
                    &1000 & 5.45 & 46.60 & 95.90 & 5.95 & 40.15 & 91.65& 5.25 & 44.70 & 96.00 & 4.95 & 36.40 & 90.00\\
                    \midrule
                    Louvain&500 & 13.95 & 44.75 & 88.30 & 11.70 & 27.60 & 62.30& 9.75 & 46.85 & 94.80 & 8.00 & 17.10 & 40.75\\
                    &1000 & 13.60 & 66.85 & 99.05 & 12.15 & 41.85 & 86.65& 7.40 & 72.75 & 99.90 & 7.05 & 25.15 & 66.50\\
                    \midrule
                    RGCR1hm&500 & 8.40 & 30.80 & 77.95 & 8.10 & 29.80 & 77.95 & 5.90 & 33.15 & 85.50 & 5.60 & 27.75 & 75.20\\
                    &1000 & 6.15 & 50.85 & 97.25 & 6.35 & 49.90 & 96.85 & 5.45 & 55.35 & 98.40 & 4.95 & 46.30 & 95.35\\
                    \midrule
					EgoCR&500 & 6.30 & 34.30 & 84.05 & 6.20 & 29.80 & 77.60& 4.70 & 39.10 & 89.95 & 4.65 & 25.70 & 71.70\\
                    &1000 & 5.90 & 56.25 & 98.35 & 5.10 & 47.20 & 96.75& 5.05 & 64.70 & 99.45 & 5.10 & 45.20 & 94.65\\
					\bottomrule
					\bottomrule
				\end{tabular}}
\end{table}

\section{Empirical Applications}\label{sec:real}
In this section, we evaluate the performance of our proposed method in two empirical applications. For comparison, we also include the results of alternative clustering methods \texttt{CR}, \texttt{CGD}, \texttt{$3$-net}, \texttt{LinkedIn}, \texttt{Louvain}, and \texttt{RGCR1hm} considered in Section~\ref{sec:numerical}.

\subsection{Application to the Sinanet Dataset}

The Sinanet dataset, collected by \cite{jia2017node}, is a microblog user relationship network extracted from the sina-microblog platform. The data were constructed by first selecting $100$ VIP users, who purchase the platform’s VIP service and are typically more active, from ten major forums, including finance and economics, literature and arts, fashion and vogue, current events and politics, sports, science and technology, entertainment, parenting and education, public welfare, and normal life. The followees of these VIP users, together with their published microblogs, were then extracted. Since we focus on undirected graphs in this paper, we treat the ties between users as undirected. The resulting network contains $n=3467$ users with an average degree of $17$. This dataset is available at \url{https://github.com/smileyan448/Sinanet}.
We also consider the setting with unobserved confounders $Z$, which is not available at either the experiment design or analysis. We let $Z$ be 
the interest distributions of each user in the ten forums, which was obtained by the Latent Dirichlet Allocation (LDA) topic model.

To apply our experimental procedure on the Sinanet network, we first construct ego-clusters using our proposed ego-clustering algorithm. Given the network structure and resulting clusters, treatments are randomly assigned at the cluster level. 
The outcomes are generated according to the following model,
$$
Y_i=\alpha+\beta T_i+\gamma\rho_i+\eta^\T Z+\epsilon_i,
$$
where $\alpha=2$, $\beta=2.5$, $\gamma=5$, $\eta=(0.6, 1.1, 3.5, 2, 1.6, 1.5, 0.8, 2.5, 3, 1.5)$, and the error terms are independently drawn from a standard normal distribution. We use model \eqref{equ:model} as the working model to estimate $\tau$ and $\gamma$. Based on $2000$ replications, the biases, standard deviations, and root mean squared errors of estimators $\hat{\tau}$ and $\hat{\gamma}$ are summarized in Table~\ref{tab:sina}. The column $K_n$ reports the number of clusters generated by each method.

\begin{table}[htpb]
	\centering
	\caption{Biases (Bias), standard deviations (SD), and root mean squared errors (RMSE) of $\hat{\tau}$ and $\hat{\gamma}$ under different clustering methods for the Sinanet dataset. Lowest RMSEs are marked in boldface.  \label{tab:sina}}
	\scalebox{0.9}{
		\begin{tabular}{llrrrrrr}
			\toprule
			\toprule
			&&\multicolumn{3}{c}{$\hat{\tau}$} & \multicolumn{3}{c}{$\hat{\gamma}$}\\
			\cmidrule(lr){3-5}\cmidrule(lr){6-8}
			Method & $K_n$ & Bias & SD & RMSE & Bias & SD & RMSE\\
			\midrule
			CR & 3467 & 0.000 & 0.115 & 0.115 & 0.001 & 0.107 & 0.107\\
			CGD & - & 0.067 & 0.462 & 0.466 & -0.017 & 0.295 & 0.295\\
			$3$-net & 59 & 0.001 & 0.104 & 0.104 & 0.002 & 0.105 & 0.105\\
			LinkedIn & 1661 & -0.001 & 0.100 & 0.100 & 0.000 & 0.102 & 0.102\\
			Louvain & 10 & -0.001 & 0.143 & 0.143 & 0.004 & 0.368 & 0.368\\
			RGCR1hm & 775 & -0.002 & 0.099 & 0.099 & -0.001 & 0.089 & 0.089\\
			EgoCR & 2005 & 0.005 & 0.094 & \textbf{0.094} & 0.005 & 0.087 & \textbf{0.088}\\
			\bottomrule
			\bottomrule
	\end{tabular}}
\end{table}

Both \texttt{LinkedIn} and \texttt{EgoCR} formulate a large number of ego-clusters and show improved efficiency compared with \texttt{Louvain} and \texttt{$3$-net}, suggesting the local ego-cluster structure is well-suited to this network. As reported in Table~\ref{tab:sina}, the biases of the two estimators remain close to zero under all methods and our proposed method EgoCR gives the smallest standard deviations and RMSEs of $\hat{\tau}$ and $\hat{\gamma}$.

\subsection{Application to a Social Network Field Experiment}

We consider a social network field experiment, which was conducted in $56$ middle schools in New Jersey, USA \citep{Paluck2016}. The experiment examined the impact of an anti-conflict intervention on students' attitudes toward conflict-related behaviors, and how these effects propagated through their social networks. At the beginning of the 2012-2013 school year, students' social network was recorded by asking them to list up to ten peers they spent time with in the last few weeks. 
In the study, half of $56$ schools were randomly selected to host the anti-conflict program. Within each selected school, a subset of students was identified as eligible for treatment, and then half of these eligible students were block-randomized into treatment based on student characteristics. The treated students attended the anti-conflict program bi-weekly during the school year. The outcome variable is the self-reported behavior on wearing a wristband, which was a reward for engaging in anti-conflict behaviors. Further details of the experiment are provided in \cite{Paluck2016}, and the dataset is available at \url{https://www.icpsr.umich.edu/web/civicleads/studies/37070/versions/V2}.

We use the observed network, covariates, and fitted parameters from the original experiment to construct a simulation study with different experimental designs.
Our simulation focuses on data from $5$ treated schools randomly selected from the original experiment, comprising $n=1214$ students. To ensure the adjacency matrix is symmetric, any two students are connected if either one reports the other as a friend. The average degree of the network is $11$. Figure~\ref{fig:plotschool} plots the network with grades distinguished by four colors. It is seen that there are no edges between different schools, and the friendships are more prevalent among students in the same grade within the same school. Using data from these schools, we first fit the following additive outcome model,
$$
Y_i=\alpha+\beta T_i+\gamma\rho_i+\eta_1 \text{GENDER}_i+\eta_2 \text{ETHW}_i+\eta_3^\T(\text{SCHOOL}_i \times \text{GRADE}_i)+\epsilon_i,
$$
where GENDER denotes student gender, ETHW is an indicator for white ethnicity, and SCHOOL and GRADE are indicator variables representing the school and grade of students, respectively. To implement our experimental procedure, we first construct ego-clusters using our algorithm and then perform ego-level randomization to assign treatments. After assignments, outcomes are generated according to the fitted model. Specifically, the true values of the global treatment effect and the spillover effect are $\tau=0.27$ and $\gamma=0.21$, respectively. Based on $2000$ replications, the biases, standard deviations and root mean squared errors of estimators $\hat{\tau}$ and $\hat{\gamma}$ are provided in Table~\ref{tab:school}. The column $K_n$ reports the number of clusters generated by each method.

\begin{figure*}[!t]
	\centering
	\includegraphics[width=0.4\textwidth]{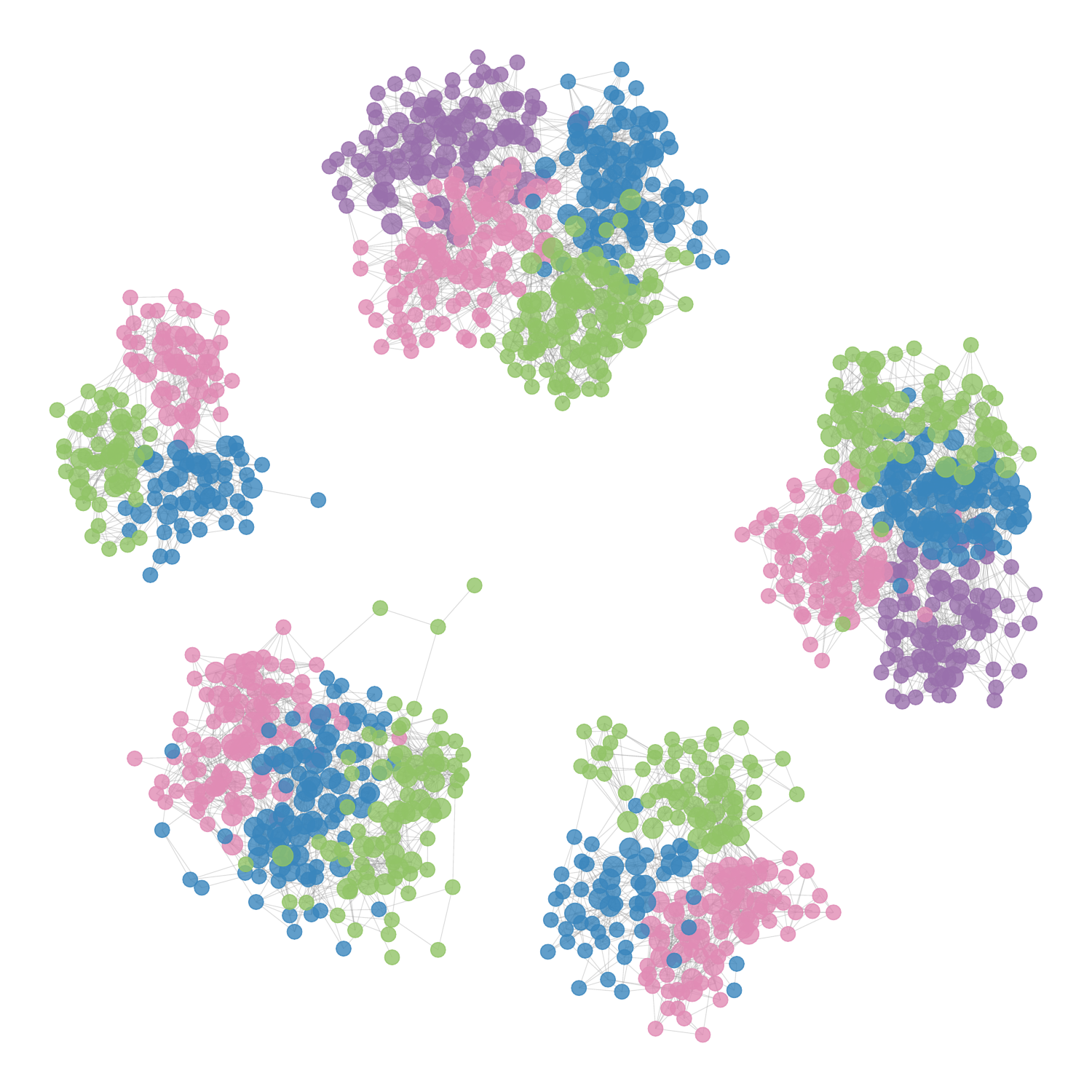}
	\caption{The network of selected $5$ schools with grades distinguished by four colors. \label{fig:plotschool}}
\end{figure*}

	\begin{table}[!t]
		\centering
		\caption{Biases, standard deviations (SD), and root mean squared errors (RMSE) of $\hat{\tau}$ and $\hat{\gamma}$ under different clustering methods for school network data. Lowest RMSEs are marked in boldface. \label{tab:school}}
		\scalebox{0.9}{
			\begin{tabular}{llrrrrrr}
				\toprule
				\toprule
				&&\multicolumn{3}{c}{$\hat{\tau}$} & \multicolumn{3}{c}{$\hat{\gamma}$}\\
				\cmidrule(lr){3-5}\cmidrule(lr){6-8}
				Method & $K_n$ & Bias & SD & RMSE & Bias & SD & RMSE\\
				\midrule
				CR & 1214 & -0.003 & 0.100 & 0.100 & -0.004 & 0.094 & 0.094\\
				CGD & - & 0.001 & 0.178 & 0.178 & 0.000 & 0.206 & 0.206\\
				CausalC & 11 & -0.003 & 0.080 & 0.080 & -0.007 & 0.210 & 0.211\\
				$3$-net & 63 & 0.003 & 0.075 & 0.075 & 0.004 & 0.090 & 0.090\\
				LinkedIn & 292 & 0.001 & 0.086 & 0.086 & 0.000 & 0.091 & 0.091\\
				Louvain & 12 & 0.001 & 0.078 & 0.078 & -0.009 & 0.216 & 0.217\\
				RGCR1hm & 298 & 0.001 & 0.074 & 0.074 & 0.002 & 0.079 & \textbf{0.079}\\
				EgoCR & 239 & -0.001 & 0.070 & \textbf{0.070} & -0.001 & 0.080 & 0.080\\
				\bottomrule
				\bottomrule
		\end{tabular}}
	\end{table}

In Table~\ref{tab:school}, the biases of $\hat{\tau}$ and $\hat{\gamma}$ are negligible under all methods. Cluster-based approaches yield smaller standard deviations for $\hat{\tau}$ compared to \texttt{CR}, showing the benefit of cluster randomization when estimating the global treatment effect in the school network. \texttt{CausalC} and \texttt{Louvain} both construct a small number of clusters with relatively low interference, which is unfavorable for estimating $\gamma$. Consequently, their standard deviations for $\hat{\gamma}$ are larger, even exceeding those under \texttt{CR}, as shown in Table~\ref{tab:school}.

\section{Conclusion}\label{sec:conclu}
This paper develops a statistical framework for estimating the global treatment effect and spillover effect in network experiments by employing the ego-cluster design and a theory guided ego-clustering algorithm. We consider a super-population framework, where the network data are viewed as a sample from an underlying population and can be obtained through the common snowball sampling scheme. Our procedure can be used to estimate both the global treatment effect and the spillover effect within a single experiment. To our knowledge, this is the first theoretical analysis of the ego-cluster design.

Building on the results of this paper, several directions for future research are worth exploring. First, the framework could be extended to handle directed networks, where units have distinct in- and out-neighbors. In such settings, the ego-cluster framework and related assumptions need to be reformulated to accommodate the asymmetric relationships. 
Second, given that more balanced allocation over covariates often improves statistical efficiency \citep{hu2012asymptotic,Bugni2018inference,ma2022new}, it can be useful to incorporate adaptive randomization into the ego-cluster design to improve covariate balance \citep{zhou2024adaptive, Liu2024}. Finally, models involving more general types of interference could be considered \citep{Aronow2017,Leung2022causal,gao2023causal}. We leave these to future research. 

\bibliographystyle{apalike}
\begingroup
\baselineskip=17.25pt
\bibliography{refs}
\endgroup

\end{document}